\documentclass[manuscript]{aastex}
\usepackage{subfigure}
\usepackage{graphicx}
\usepackage{color}

\received{2014 February 15}
\accepted{2014 February 16}

\shorttitle{Can Charge Exchange Explain Anomalous Soft X-ray Emission in the
    Cygnus Loop?}
\shortauthors{R. S. Cumbee et al.}

\begin{document}

\title{Can Charge Exchange Explain Anomalous Soft X-ray Emission in the
    Cygnus Loop?}

\author{R. S. Cumbee\altaffilmark{1}, D. B. Henley\altaffilmark{1}, P. C. Stancil\altaffilmark{1}, R. L. Shelton\altaffilmark{1}, 
J. L. Nolte\altaffilmark{1,2,3}, Y. Wu\altaffilmark{1,4}, and D. R. Schultz\altaffilmark{5} }

\altaffiltext{1}{Department of Physics and Astronomy and the Center for Simulational Physics, University of Georgia, Athens, GA 30602, USA}
\altaffiltext{2}{Department of Physics, Western Michigan University, Kalamazoo, MI 49008, USA}
\altaffiltext{3}{Centre for Theoretical Atomic, Molecular and Optical Physics, Queen's University Belfast, Belfast BT7 1NN, UK}
\altaffiltext{4}{Institute of Applied Physics and Computational Mathematics,  Beijing, PR China}
\altaffiltext{5}{Department of Physics, University of North Texas, Denton, Texas 76203, USA}

\begin{abstract}
Recent X-ray studies have shown that supernova shock models are unable to satisfactorily explain X-ray emission
in the rim of the Cygnus Loop. In an attempt to account for this ``anomalously'' enhanced X-ray flux, we fit the 
region with a model including theoretical charge exchange (CX) data
along with shock and background X-ray models. The model includes the CX collisions of O$^{8+}$, O$^{7+}$,
N$^{7+}$, N$^{6+}$, C$^{6+}$, and C$^{5+}$ with H with an energy of 
1 keV/u (438 km/s). The observations reveal a strong emission feature
near 0.7 keV that cannot fully be accounted for by a shock model, nor the 
current CX data. Inclusion of CX, specifically O$^{7+}$ + H,
does provide for a statistically significant improvement over 
a pure shock model.

\end{abstract}

\keywords{atomic processes, line: formations, ISM: supernova remnants, X-rays: general }

\section{Introduction}

Charge exchange (CX) is a fundamental atomic collision process to which the X-ray spectra
of various astrophysical environments have been attributed  \citep[see, for example,][]{den10,kra04}.
In this process (also referred to as 
charge transfer or electron capture), X-ray emission is a consequence of a highly charged ion capturing an electron from another atom,
\begin{equation}
A^{q+} + B \to A^{(q-1)+}(nl~^{2S+1}L) + B^+,
\label{eq1}
\end{equation}
\noindent to form a highly excited, high charge state ion 
$A^{(q-1)+}(nl~^{2S+1}L)$. As the electron cascades down to the lowest energy level,
X-rays are emitted. To accurately model charge exchange emission spectra, it is essential to include the effects of many
ionization stages, $q$, and donor (neutral target) species, $B$, for a given element (projectile) $A$. 
The process has been found to play a role in X-ray emission from
the atmospheres of Mars and other planets \citep{den02},  
 from starburst galaxies \citep{liu11}, and from extragalactic 
cooling flows \citep{fab11,lal04}.

From \textit{Suzaku} observations, \citet{kat11} suggested that the outer rim of the Cygnus Loop may be a prime example of extra-solar CX X-ray emission.
In their comparisons between the observed spectra and a model that included shocked gas and an X-ray background, they were unable to fully reproduce
the spectra, which suggested that CX may be missing from their model. They then adjusted their model by including lines with relevant energies and widths but arbitrary intensities
to mimic CX emission spectra. In order to improve upon the \citet{kat11} approach, we replace their CX-inspired lines with 
theoretical CX data to model X-ray emission in the rim of the Cygnus Loop.

\section{Observations and Data Reduction}

Here, we analyze a \textit{Suzaku} observation of one of the fields studied along the rim of the Cygnus Loop by \citet{kat11}, 
labeled NE4 in their Figure 1. NE4 includes part of the northeastern rim of
the Cygnus Loop and is a region believed to have a significant neutral H fraction. The pointing direction was $(\alpha, \delta) = (20^\mathrm{h}54^\mathrm{m}00\fs12,
+32\degr22\arcmin08\farcs4)$. The observation was carried out on 2005 Nov 30 and given \textit{Suzaku} observation ID 500023010.

\begin{figure}
\includegraphics[scale=.42]{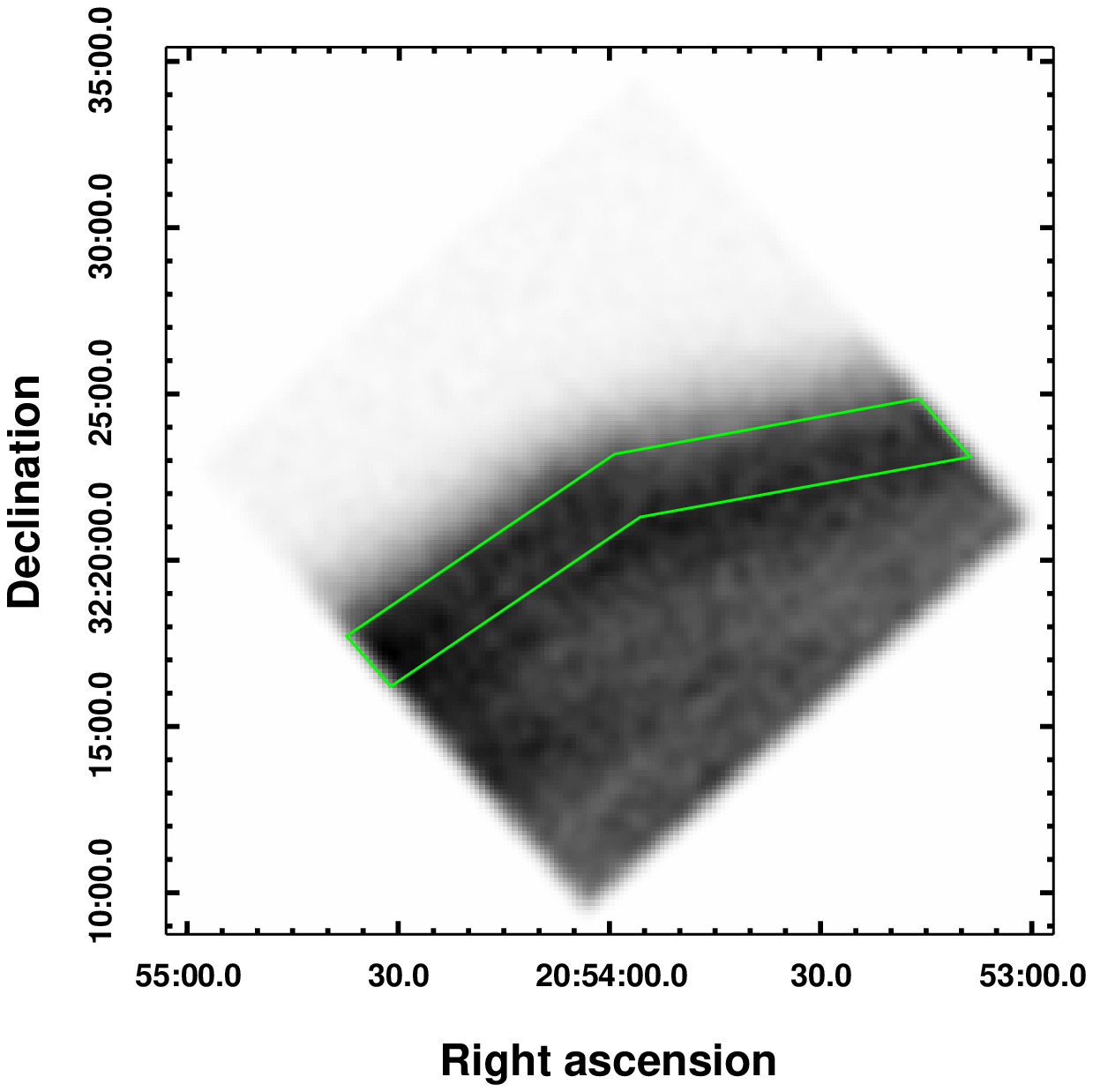}
\includegraphics[scale=.22]{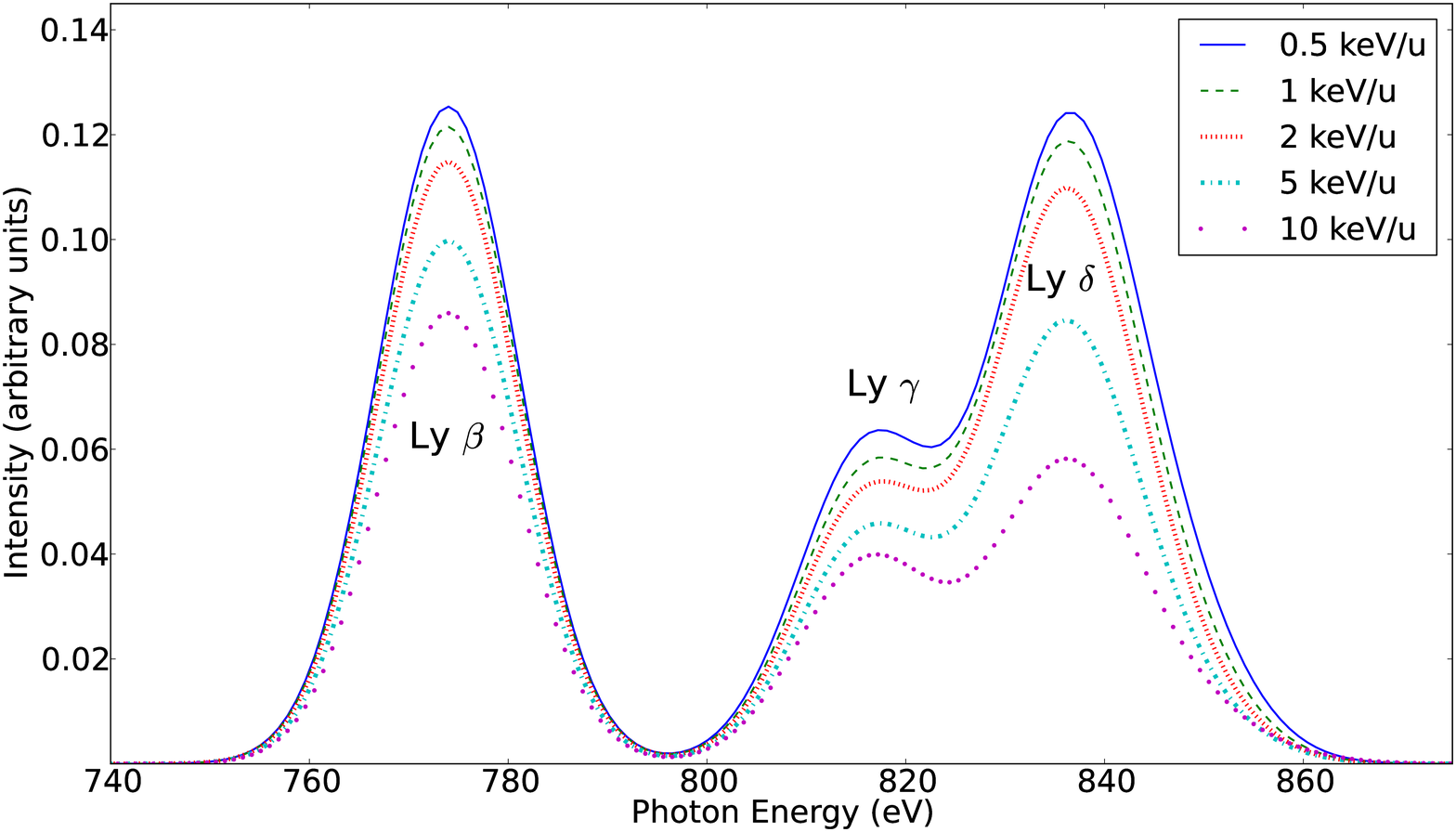}
\put(-480,120){(a)}
\put(-260,120){(b)}
\caption{a) Smoothed 0.3--2.0~keV XIS1 image of the NE4 field of the Cygnus Loop. The image has not
  been background-subtracted nor flat-fielded. The green polygon shows the region from which we
  extracted the spectrum of the Cygnus Loop rim.
b) X-ray emission spectrum including Ly$\beta$ and higher lines for the collision of fully ionized O with neutral H for a 
collisional energy of 0.5, 1, 2, 5, and 10 keV/u (309, 438, 619, 979, and 1384 km/s, respectively). The lines are convolved with a FWHM of 10 eV.
For each collisional energy, the Ly$\beta$ and higher lines are normalized to the Ly$\alpha$ line.
Cross sections are from \citet{jan93}.}.\label{fig1}
\end{figure}

We reduced the XIS data and extracted X-ray spectra using version 6.12 of
HEASoft,\footnote{http://heasarc.gsfc.nasa.gov/lheasoft/} using the calibration data that was
current as of 2013 Mar 23. We first ran the \texttt{aepipeline} script, to reprocess the events list
using the most recent calibration data. This script also screens the data -- we applied the standard
screening recommended in the \textit{Suzaku} ABC
Guide.\footnote{http://heasarc.gsfc.nasa.gov/docs/suzaku/analysis/abc/abc.html} Then, for each XIS
camera, we combined the data taken in the $3 \times 3$ and $5 \times 5$
modes. Figure~\ref{fig1}(a) shows the resulting 0.3--2.0~keV X-ray image from the XIS1 camera.

From each of the four XIS cameras we extracted an X-ray spectrum of the bright rim of the Cygnus
Loop.  We extracted these spectra from a $\sim$2\arcmin\ wide strip positioned by eye along the rim
(the green polygon in Figure~\ref{fig1}(a)). We grouped each spectrum such that there were at
least 25 counts per spectral bin. For each of these spectra we generated a corresponding non-X-ray
background (NXB) spectrum, to be subtracted from the source spectrum in the subsequent spectral
analysis. For this purpose we used the \texttt{xisnxbgen} tool \citep{tawa08}, which constructs NXB
spectra using a database of \textit{Suzaku} observations of the nightside of Earth. We also generated the
response files for each spectrum, using \texttt{xisrmfgen} to calculate the redistribution matrix
files (RMFs) and \texttt{xissimarfgen} \citep{ishisaki07} to calculate ancillary response files
(ARFs), assuming a uniform extended source. Note that the latter tool takes into account the
temporally and spatially varying contamination on the XIS cameras' optical blocking filters.

We combined the spectra from the three front-illuminated (FI) cameras (XIS0, XIS2, and XIS3) into a
single spectrum (with corresponding NXB spectrum and response files) using \texttt{addascaspec}.
In the following spectral analysis, we fitted our models to this combined FI spectrum and to the
spectrum from the back-illuminated (BI) XIS1 camera simultaneously.

\section{Charge Exchange Induced X-ray Emission Cascade Model}

In order to produce the CX-induced portion of the spectra to be used in our model of the Cygnus Loop,
we applied a low density, steady-state radiative cascade model as described in \citet{rig02}. In this model, the
initial state populations are proportional to the quantum-state-resolved CX cross sections. As the 
electron cascades down to the lowest energy level obeying quantum mechanical selection rules, 
photons are emitted, including X-rays. 
Our model includes X-ray emission spectra produced using cross sections from the best available
sources for the ions O$^{8+}$ \citep{jan93}, O$^{7+}$ \citep{nol12b}, 
N$^{7+}$ \citep{har98}, N$^{6+}$ \citep{wu11}, C$^{6+}$ \citep{jan93}, and C$^{5+}$ \citep{nol12}
colliding with atomic H with a collision energy of 1 keV/u (438 km/s).

We considered only ion interactions with neutral H as CX cross section data for He are sparse.
However, as the H/He elemental abundance ratio is $\sim$10, neglecting CX reactions involving He should not adversely affect our results. 
Varying the energy and including He as a target atom along with explicit CX data for Ne, Mg, Fe,
and other ions are improvements that will be made in future work. In Figure~\ref{fig1}(b),
example \ion{O}{8} X-ray emission spectra for O$^{8+}$ colliding with H for five collisional energies are given. 
While \citet{gha01} derived a shock speed of 270-350 km/s from the shock's H$\alpha$ profile, as
Figure~\ref{fig1}(b) shows, there is little variation in the intensities of the K$\beta$ and higher lines between the 
collisional energies of 0.5 and 2 keV/u. As data from \citet{jan93} are most reliable above 1 keV/u, we chose to use that energy in the current model.

\section{Analysis and Results}

\begin{figure}
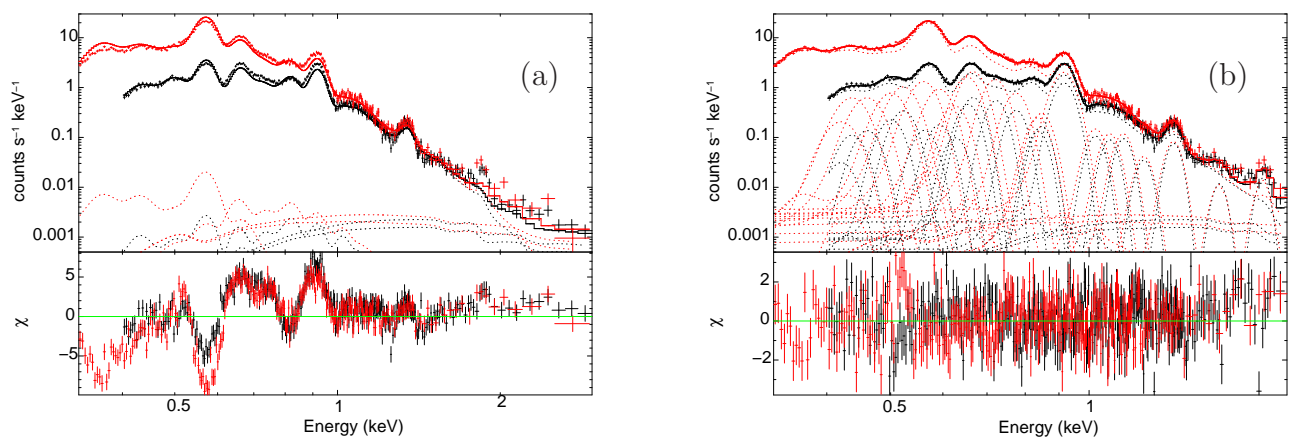

\includegraphics[scale=.32, angle = 270]{fig2a.ps}
\includegraphics[scale=.32, angle = 270]{fig2b.ps}
\put(-320,-40){(a)}
\put(-60,-40){(b)}
\caption{(a) Spectrum (crosses) from NE4's outermost rim of the Cygnus Loop as described in Figure 1 of \citet{kat11} along with the best-fit
model (lines) including the Absorbed VpShock model. BI spectrum (red), FI spectrum (black). Residuals are shown in the lower panel with a reduced $\chi^2$ of 8.9 with 592 d.o.f.
(b) Same as panel (a), but with the Absorbed VpShock model and CX-inspired lines whose intensities are
independently varied in the fit. This figure can be compared to Figure 6 from \citet{kat11}. Residuals show a reduced $\chi^2$ = 1.36 with 533 d.o.f.}\label{fig2}
\end{figure}

Our first step in analyzing the rim spectra was to determine how well a model spectrum without CX fit the observations.
Our model spectra without CX was patterned on that used by \citet{kat11}. We used 
 XSPEC version 12.8.1 \citep{ARN96} and applied an absorbed VpShock model with a hydrogen column density for the 
 intervening material of the Cygnus Loop of 3$\times10^{20}$cm$^{-2}$ \citep{kos10}. In addition, two APEC thermal plasma models and two broken power laws for the X-ray background 
with the parameters described by \citet{kat11} and \citet{yos09} were used with a total Galactic hydrogen column
density of 1.7$\times 10^{21}$cm$^{-2}$ for the direction of the Cygnus Loop as described in the Leiden/Argentine/Bonn 
Survey of Galactic HI \citep{kal05}.
As Figure~\ref{fig2}(a) shows, 
we were not able to adequately fit the spectrum, particularly below 1 keV. Our model resulted in a
reduced $\chi^2$ of 8.9 with 592 degrees of freedom (d.o.f.), compared to a reduced $\chi^2$ of 2.21 obtained by \citet{kat11}, even though we followed similar steps in extracting the data (see \S2)
and applied the same model.

Finding their version of this model unsatisfactory, \citet{kat11} went on to construct a thermal plus CX-inspired model in which the thermal components were drawn from fits to a nearby
part of the rim that was not thought to experience CX (their NE2 field) and the CX contribution was modeled as 22 independent emission lines. 
To compare directly to Figure 6 of \citet{kat11}, we introduced 35 lines to the previous model,
in a similar manner in which they introduced 22 lines. Following their steps, we adopted the VpShock parameters for NE2 given in their Table 2.
This included an electron temperature of 0.21 keV, and C, N, O, Ne, Mg, Si, and Fe elemental
abundances of 0.17, 0.07, 0.08, 0.15, 0.10, 0.10, and 0.14 times the respective solar value.
However, unlike Katsuda et al., we resolve all multiplets, hence the additional 13 lines.
We assumed that the lines' intrinsic widths are much smaller
than the instrumental line width. At this stage, the normalization of each line was allowed to vary independently as shown in Figure~\ref{fig2}(b).
While this resulted in a much improved reduced $\chi^2$ of 1.36 with 533 d.o.f., the
relative line intensities of a given ion are unconstrained by laboratory data and can lead to unphysical line ratios. 
As column 3 in Table~\ref{tab1} shows, for cases such as \ion{C}{6} where Ly$\alpha$ is essentially zero, the
Ly$\beta$/Ly$\alpha$ line ratio is much greater than 1.
The higher-order CX emission lines have intensities which are usually smaller than the K$\alpha$ feature. 
An example is shown in the case of  O$^{8+}$ colliding with H in Figure~\ref{fig1}(b) in which
the Ly$\beta$/Ly$\alpha$ line ratio is $\sim$0.1.

\begin{figure}
\includegraphics[scale=.62, angle = 270]{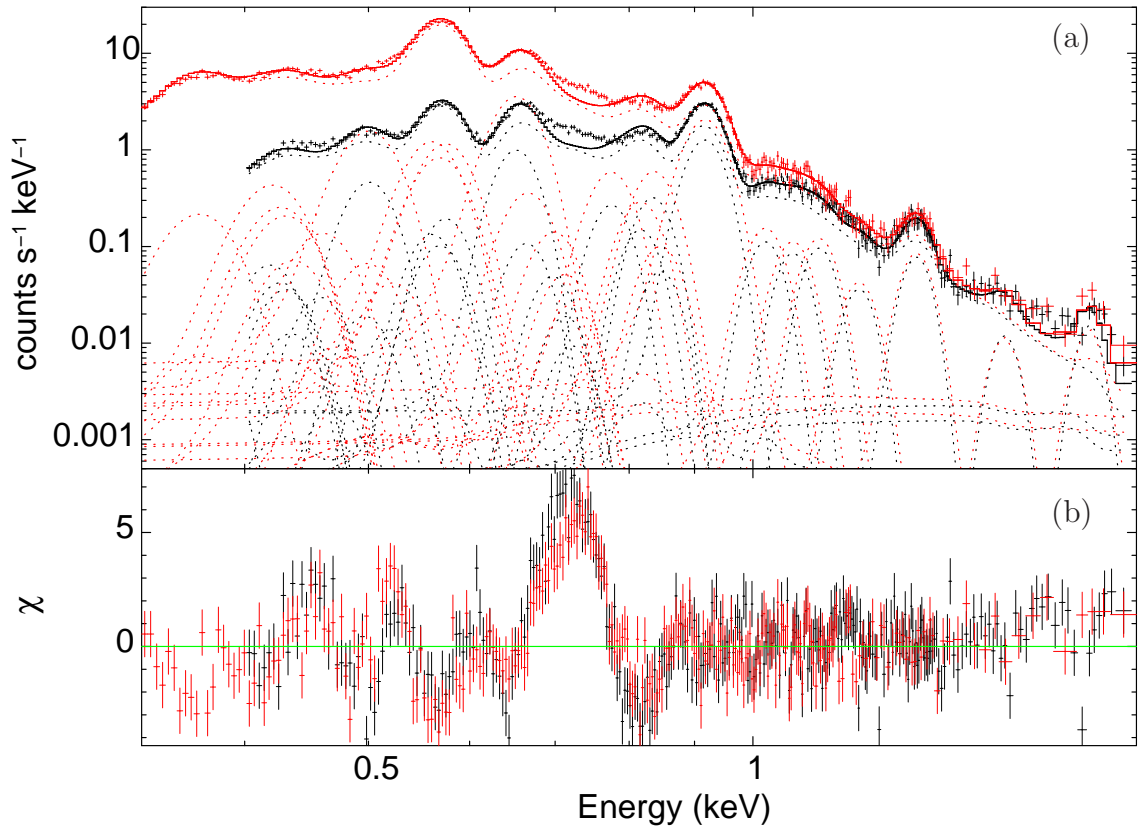}
\put(-90,-40){(a)}
\put(-90,-220){(b)}
\caption{a) Same as Figure \ref{fig2}(b), but with the Absorbed VpShock model and theoretical CX data. The normalizations
for K$\beta$ and higher are tied to the K$\alpha$ resonant lines for each ion to more accurately depict CX emission spectra.
b) Reduced $\chi^2$ = 3.95 with 555 d.o.f. primarily due to the poor fit near 0.7 keV.}\label{fig3}
\end{figure}

In order to incorporate realistic CX spectra, we then tied the normalizations of the K$\beta$ and higher lines to their
respective K$\alpha$ resonant line, as shown in Figure~\ref{fig3}, based on the CX cascade spectra described above.
The normalizations are given in column 4 in Table~\ref{tab1}. This 
resulted in a reduced $\chi^2$ of 3.95 with 555 d.o.f. 
While the fit quality is reduced relative to the unconstrained spectra, 
the CX line ratios are physically reasonable, being obtained from theoretical CX cross sections.

\subsection{The 0.7 keV Region}
\begin{figure}
\includegraphics[scale=.62, angle = 270]{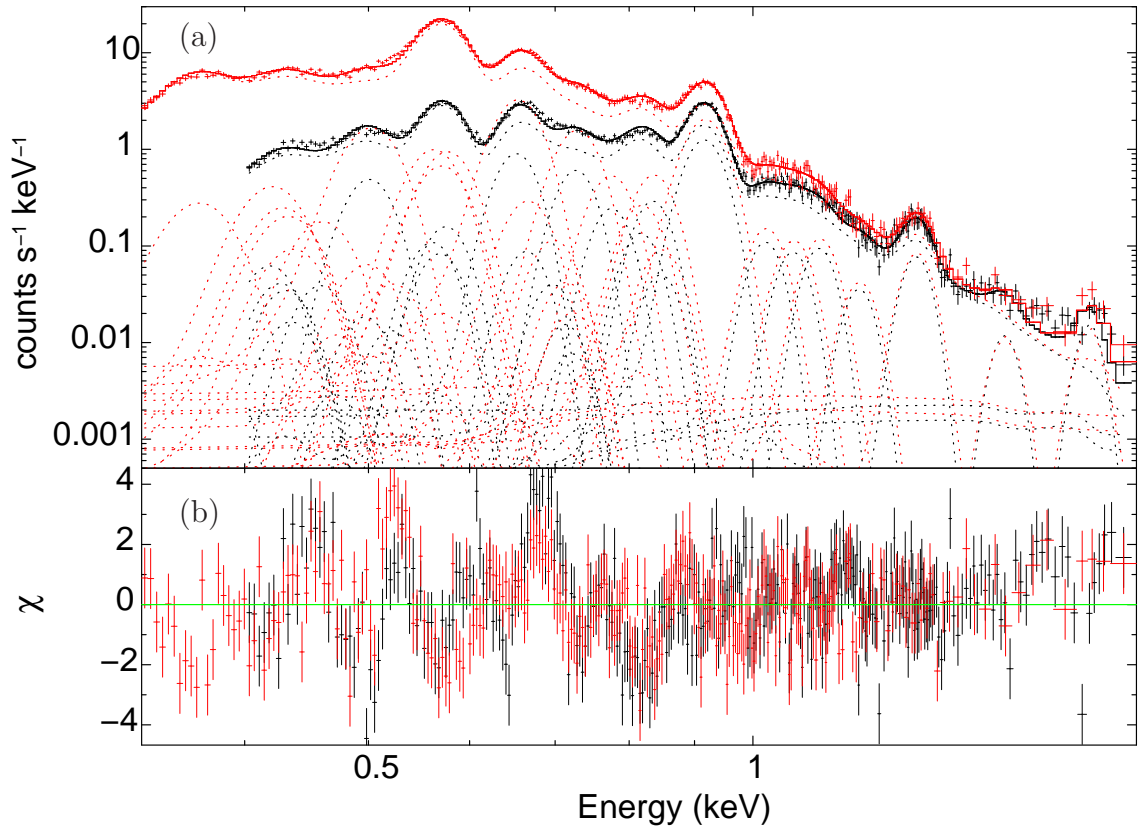}
\put(-420,-40){(a)}
\put(-420,-220){(b)}
\caption{a) Same as Figure \ref{fig3}, but with the addition of Fe L-shell lines at 0.726 and 0.822
keV. b) Reduced $\chi^2$ = 1.83 with 553 d.o.f.}\label{fig4}
\end{figure}

\citet{kat11} found that their shock + background model was not as bright around 0.7 keV as the observed spectrum. Hence they
suggested that the 0.7 keV band might be due to \ion{Fe}{17}/\ion{Fe}{16} L-shell emission 
and/or O K-shell CX. They ultimately discarded the Fe contribution in their final models.
Our CX + shock + background model is also unable to fully account for the observed 0.7 keV emission. However, as we included realistic
oxygen CX, we note that oxygen CX emission does contribute to the spectrum, but it alone cannot fully explain the feature. 
Since Fe$^{16+}$ and Fe$^{17+}$ + H CX line ratios are not available, we modeled these spectral features with two unconstrained lines at 0.726 keV and 
0.822 keV, following the original suggestion of \citet{kat11}. As Figure~\ref{fig4} shows, the spectrum fit is improved, resulting
in a reduced $\chi^2$ of 1.83 with 553 d.o.f. Our model,  as described in Column 5 of Table~\ref{tab1}, found zero intensity for the second line at 0.822 keV in contrast with expectations
based on Fe L-shell line ratios found in thermal models and observations 
in other X-ray emitting regions (see Katsuda et al. and references therein).

\section{Discussion}
With the advent of high-resolution X-ray detectors, apparently anomalous X-ray 
features have been observed in a variety of environments.
When the spectra cannot be explained by thermal electron impact or shock models, for example, CX
is often invoked, but typically without realistic CX data.
In this Letter, we have incorporated realistic CX X-ray emission models in an attempt to test the suggestion 
of \citet{kat11} that the anomalous X-ray emission near 0.7 keV in the Cygnus Loop 
supernova remnant is due to CX. We find the following major points:
 (i) charge exchange likely contributes to the X-ray spectrum of the Cygnus Loop from 0.3 - 2 keV;
 (ii) when obtaining CX-induced emission line intensities, lines with unconstrained normalizations may lead to unphysical line ratios;
 (iii) as suggested by \citet{kat11}, high order \ion{O}{7} emission lines due to O$^{7+}$ + H CX contributes to
       the 0.7 keV spectral features, however the \ion{O}{7} line intensities constrained by CX emission models 
       are insufficient to fully account for the 0.7 keV X-ray flux; and
 (iv) the missing flux at 0.7 keV may be due to \ion{Fe}{17} CX and/or emission from a species and/or mechanism not included in the current models.

In conclusion, using accurate CX X-ray emission models for C, N, and O ions, we confirm that charge exchange may contribute to the X-ray emission  
observed from a supernova remnant, the Cygnus Loop. However, detailed models to accurately
simulate the spectrum to extract physical conditions of the local environment require additional CX studies, particularly for other elements and charge states.

\acknowledgments
This work was partially supported by NASA grants NNX09AC46G and NNG09WF24I.
D.B.H. acknowledges funding from NASA Astrophysics Data Analysis Program
grant NNX12AI56G and NASA/SAO \textit{Chandra} grant AR2-13017X.

\begin{deluxetable}{c c c  c c}
\tabletypesize{\scriptsize}

\tablecaption{\label{tab1} Model Line Normalizations}
\tablewidth{0pt}
\tablehead{
\multicolumn{1}{c}{Line} & 
\multicolumn{1}{c}{Energy} &
\multicolumn{1}{c}{CX-Inspired (Fig~\ref{fig2}(b))} & 
\multicolumn{1}{c}{CX (Fig~\ref{fig3})} & 
\multicolumn{1}{c}{CX with Fe (Fig~\ref{fig4})}\\
\multicolumn{1}{c}{} & 
\multicolumn{1}{c}{(keV)} & 
\multicolumn{1}{c}{(L.U.)}& 
\multicolumn{1}{c}{(L.U.)}& 
\multicolumn{1}{c}{(L.U.)}\\
\multicolumn{1}{c}{(1)} & 
\multicolumn{1}{c}{(2)} & 
\multicolumn{1}{c}{(3)}& 
\multicolumn{1}{c}{(4)}& 
\multicolumn{1}{c}{(5)}
}
\startdata
C$^{5+}$ +H  K$\alpha$     & 0.3079 &	 2.8$\times 10 ^{-13}$ $^{+3000.4}_{-0.0}$ 	& 6.9$\times 10 ^{-14}$	$^{+52.6}_{-0.0}$ & 7.0$\times 10 ^{-14}$	$^{+68.1}_{-0.0}$	\\
C$^{5+}$ +H  K$\beta$      & 0.3545 &	 1.0$\times 10 ^{-13}$ $^{+130.2}_{-0.0}$ 	& 2.5$\times 10 ^{-14}$ 	& 2.5$\times 10 ^{-14}$ \\
C$^{5+}$ +H  K$\gamma$     & 0.3709 &	 66.2 $^{+25.3}_{-66.2}$ 	& 4.3$\times 10 ^{-15}$ 	& 4.3$\times 10 ^{-15}$ \\
C$^{6+}$ +H  Ly$\alpha$    & 0.3673 &	 1.0$\times 10 ^{-13}$ $^{+120.1}_{-0.0}$ 	& 73.27	$^{+98.41}_{-70.05}$ & 128.7	$^{+130.4}_{-80.6}$	\\
C$^{6+}$ +H  Ly$\beta$     & 0.4353 &	 193.9 $^{+2564.2}_{-0.0}$ 	& 10.83 	& 19.02 \\
C$^{6+}$ +H  Ly$\gamma$    & 0.4591 &	 165.2 $^{+55.8}_{-91.7}$ 	& 16.75 	& 29.41 \\
C$^{6+}$ +H  Ly$\delta$    & 0.4701 &	 45.60 $^{+102.07}_{-39.67}$ 	& 1.026 	& 1.801 \\
N$^{6+}$ +H  K$\alpha$  f  & 0.4197 &	 39.54 $^{+158.24}_{-11.7}$ 	& 265.1	$^{+42.0}_{-20.2}$ & 246.7	$^{+27.6}_{-44.2}$	\\
N$^{6+}$ +H  K$\alpha$  i  & 0.4261 &	 49.89 $^{+2702.12}_{-4.09}$ 	& 143.2 	& 133.3 \\
N$^{6+}$ +H  K$\alpha$  r  & 0.4307 &	 115.3 $^{+2669.3}_{-2.3}$ 	& 111.2 	& 103.5 \\
N$^{6+}$ +H  K$\beta$      & 0.4979 &	 7.3$\times 10 ^{-12}$ $^{+31.24}_{-0.0}$ 	& 20.32 	& 18.91 \\
N$^{6+}$ +H  K$\gamma$     & 0.5215 &	 217.8 $^{+35.51}_{-65.17}$ 	& 10.98 	& 10.22 \\
O$^{7+}$ +H  K$\alpha$ f   & 0.5609 &	 471.8 $^{+70.3}_{-119.5}$ 	& 338.5	$^{+18.2}_{-23.7}$ & 282.0	$^{+12.8}_{-23.5}$	\\
O$^{7+}$ +H  K$\alpha$ i   & 0.5684 &	 30.4 $^{+128.7}_{-30.4}$ 	& 222.9 	& 185.7 \\
O$^{7+}$ +H  K$\alpha$ r   & 0.574 &	 2.7$\times 10 ^{-14}$ $^{+109.1}_{-0.0}$ 	& 301.1 	& 250.8 \\
O$^{7+}$ +H  K$\beta$      & 0.6656 &	 200.3 $^{+78.6}_{-41.4}$ 	& 60.69 	& 50.55 \\
O$^{7+}$ +H  K$\gamma$     & 0.6978 &	 18.61 $^{+52.8}_{-1.57}$ 	& 21.83 	& 18.18 \\
O$^{7+}$ +H  K$\delta$     & 0.7127 &	 46.18 $^{+33.11}_{-3.56}$ 	& 7.659 	& 6.380 \\
O$^{7+}$ +H  K$\epsilon$   & 0.7208 &	 181.5 $^{+14.5}_{-29.6}$ 	& 3.649 	& 3.040 \\
O$^{8+}$ +H  Ly$\alpha$    & 0.653 &	 469.7 $^{+63.5}_{-12.9}$ 	& 614.2	$^{+13.9}_{-21.9}$ & 553.7	$^{+22.9}_{-29.5}$	\\
O$^{8+}$ +H  Ly$\beta$     & 0.7739 &	 101.8 $^{+15.0}_{-8.8}$ 	& 74.47 	& 67.13 \\
O$^{8+}$ +H  Ly$\gamma$    & 0.8163 &	 18.0 $^{+7.3}_{-0.7}$ 	& 33.99 	& 30.64 \\
O$^{8+}$ +H  Ly$\delta$    & 0.8358 &	 51.25 $^{+31.16}_{-15.53}$ 	& 68.38 	& 61.64 \\
O$^{8+}$ +H  Ly$\epsilon$  & 0.8465 &	 34.1 $^{+191.1}_{-34.1}$ 	& 11.30 	& 10.19 \\
N$^{7+}$ +H  Ly$\alpha$    & 0.4999 &	 214.4 $^{+17.8}_{-62.2}$ 	& 394.9	$^{+23.2}_{-26.7}$ & 420.6	$^{+31.8}_{-21.1}$	\\
N$^{7+}$ +H  Ly$\beta$     & 0.5926 &	 131.2 $^{+41.6}_{-30.0}$ 	& 40.63 	& 43.28 \\
N$^{7+}$ +H  Ly$\gamma$    & 0.625 &	 101.5 $^{+17.1}_{-25.5}$ 	& 24.10 	& 25.68 \\
N$^{7+}$ +H  Ly$\delta$    & 0.6399 &	 97.5 $^{+19.5}_{-23.5}$ 	& 77.23 	& 82.27 \\
N$^{7+}$ +H  Ly$\epsilon$  & 0.6481 & 75.0 $^{+63.1}_{-5.2}$ & 3.8$\times 10 ^{-5}$  	& 1.0$\times 10 ^{-3}$ \\
Ne IX  K$\alpha$        & 0.91 &	 257.8 $^{+12.3}_{-11.5}$ 	& 247.2	$^{+6.5}_{-6.8}$ & 251.5	$^{+6.9}_{-6.5}$	\\
Ne X  Ly$\alpha$         & 1.022 &	 14.78 $^{+1.98}_{-2.68}$ 	& 15.66	$^{+2.07}_{-2.43}$ & 15.27	$^{+2.21}_{-2.26}$	\\
Fe  L($3d\rightarrow 2p$)           & 0.726 & \nodata 	& \nodata 	& 234.8	$^{+12.6}_{-10.6}$	\\
Fe  L($3s\rightarrow 2p$)           & 0.822 & \nodata 	& \nodata 	& 2.2$\times 10 ^{-15}$	$^{+1.88}_{-0.00}$	\\
Ne   K$\beta$           & 1.074 &	 11.47 $^{+2.45}_{-2.33}$ 	& 11.39	$^{+1.94}_{-2.04}$ & 11.39	$^{+1.5}_{-2.7}$	\\
Ne   K$\gamma$          & 1.127 &	 11.00 $^{+1.72}_{-1.71}$ 	& 11.32	$^{+1.62}_{-1.56}$ & 11.46	$^{+1.11}_{-2.26}$	\\
Ne   Ly$\beta$           & 1.21 &	 3.346 $^{+0.915}_{-1.079}$ 	& 3.750	$^{+0.856}_{-1.153}$ & 3.658	$^{+0.994}_{-0.995}$	\\
Mg   K$\alpha$          & 1.343 &	 7.321 $^{+1.225}_{-0.769}$ 	& 6.849	$^{+1.118}_{-0.843}$ & 7.674	$^{+1.016}_{-0.928}$	\\
Mg  K$\beta$            & 1.571 &	 0.723 $^{+0.553}_{-0.496}$ 	& 0.807	$^{+0.482}_{-0.487}$ & 0.985	$^{+0.492}_{-0.479}$	\\
Si   K$\alpha$          & 1.833 &	 1.138 $^{+0.471}_{-0.366}$ 	& 1.182	$^{+0.244}_{-0.494}$ & 1.180	$^{+0.353}_{-0.341}$	\\
\hline
Reduced $\chi^2$         & \nodata & 1.36 with 533 d.o.f.  & 3.95 with 555 d.o.f. &  1.83 with 553 d.o.f. \\
\enddata
\tablecomments{\ Line normalizations for Figures~\ref{fig2}(b)--\ref{fig4}. L.U. (line unit) = photons cm$^{-2}$s$^{-1}$sr$^{-1}$. 
The line energies are fixed, and the lines have zero intrinsic width. In columns 4 and 5, the K$\beta$ and higher lines
 for C, N, and O are tied to their respective K$\alpha$ resonant lines (see \S3 for details).}
\end{deluxetable}

\end{document}